\documentclass[aps,amsmath,amssymb,prl,superscriptaddress,twocolumn]{revtex4-2}  

\pdfoutput=1

\usepackage{mathrsfs, amsmath, amsthm, amssymb, color, epstopdf, verbatim, enumerate, graphicx, hyperref}
\usepackage{dcolumn}   
\usepackage{bm}        
\usepackage{lineno}

\newcommand\diff{\,\mathrm{d}}

\def\sige{\sigma_{\chi,e}}

\def\nuflub8{\phi^\nu_B}
\def\nuflube7{\phi^\nu_{Be}}

\def\rhodm{\rho_{\chi}}
\def\rdm{r_{\chi}}
\def\mdm{m_{\chi}}

\def\Ldm{L_{\chi}}
\def\udm{u_{\chi}}
\def\lesim{\lesssim}

\def\vstar{v_{*}}
\def\Vstar{V_{*}}

\def\vesc{v_{\mathrm{esc}}}

\def\Rstar{R_{*}}
\def\Lstar{L_{*}}
\def\rhostar{\rho_{*}}
\def\Mstar{M_{*}}

\def\ndm{n_{\chi}}
\def\Ndm{N_{\chi}}
\def\Ldm{L_{\chi}}

\def\Pobs{P_{\mathrm{obs}}}
\def\dPobs{\dot{P}_{\mathrm{obs}}}

\def\Pthe{P_{\mathrm{the}}}

\def\dPthe{\dot{P}_{\mathrm{the}}}
\def\Msun{M_{\odot}}
\def\Rsun{R_{\odot}}
\def\Lsun{L_{\odot}}
\def\vs{v_{s}}
\def\vt{v_{t}}
\def\dP{\dot{P}}
\def\Ein{E^\mathrm{in}}

\def\Eout{E^\mathrm{out}}

\def\Enet{E^\mathrm{net}}
\def\nat{Nature\ }
\def\aap{Astron.\ Astrophys.\ }
\def\aapr{Astron.\ Astrophys.\ Rev.\ }
\def\apj{Astrophys.\ J.\ }
\def\apjl{Astrophys.\ J.\ Lett.\ }

\def\aj{Astron.\ J.\ }
\def\mnras{Mon.\ Not.\ Roy.\ Astron.\ Soc.\ }

\def\prd{Phys.\ Rev.\ D\ }
\def\prl{Phys.\ Rev.\ Lett.\ }

\def\araa{Annu.\ Rev.\ Astron.\ Astrophys.\ }
\def\jcap{J.\ Cosmol.\ Astropart.\ Phys.\ }

\def\pasp{Publications\ of\ the\ Astronomical\ Society\ of\ the\ Pacific}
\def\epjc{European\ Physical\ Journal\ C}
\def\na{New\ Astronomy\ }

\def\kpc{\,\mathrm{kpc}}
\def\km{\,\mathrm{km}}
\def\TeV{\,\mathrm{TeV}}
\def\GeV{\,\mathrm{GeV}}
\def\MeV{\,\mathrm{MeV}}

\def\MV\TeV{\,\mathrm{MV}}
\def\cm{\,\mathrm{cm}}
\def\s{\,\mathrm{s}}

\newcolumntype{p}{D{,}{\pm}{-1}}

\begin{document}
\title{Exploring the Dark Frontier: White Dwarf-Based Constraints on Light Dark Matter}

\author{Jia-Shu Niu}
\email{jsniu@sxu.edu.cn}
\thanks{corresponding author}
\affiliation{Institute of Theoretical Physics, Shanxi University, Taiyuan, 030006, China}
\affiliation{State Key Laboratory of Quantum Optics and Quantum Optics Devices, Shanxi University, Taiyuan 030006, China}
\affiliation{Collaborative Innovation Center of Extreme Optics, Shanxi University, Taiyuan 030006, China}

\date{\today}

\begin{abstract}
In the vast expanse of our galaxy, white dwarfs (WDs) are natural sentinels, capturing the enigmatic dark matter (DM) particles that incessantly traverse their interiors. These celestial bodies provide a unique vantage point for probing interactions between DM particles and their constituents-nuclei or electrons-should such interactions exist. The captured DM particles may accumulate, undergo mutual annihilation, or be evaporated by the WD's own nuclei or electrons, thereby perturbing the standard cooling sequence predicted by stellar evolution theory.
This letter reports pioneering constraints on DM-electron interactions derived from an in-depth analysis of four pulsating WDs. By leveraging the period variation rates of their pulsation modes, we delineate the following constraints: for a form factor $F(q) = 1$, in the DM mass range $20 \MeV/c^{2} \lesim \mdm \lesim 80 \MeV/c^{2}$ with a cross-section limit of $\sige \lesim 10^{-56} \cm^{2}$; for a form factor $F(q) = (\alpha m_{e})^{2}/q^{2}$, in a the DM mass range $20 \MeV/c^{2} \lesim \mdm \lesim 70 \MeV/c^{2}$ with a limit of $\sige \lesim 10^{-52} \cm^{2}$.
These newly established constraints surpass current direct detection experiments by over fifteen orders of magnitude, forging a path into the uncharted territories of the DM parameter space. This work not only advances our understanding of light dark matter-electron interactions but also exemplifies the potential of WDs as important astrophysical laboratories for probing the elusive nature of DM.
\end{abstract}

\maketitle

{\bf Motivation.}
Although dark matter (DM) is the dominant component of the matter in the Universe \cite{Plank2018}, its particle nature remains largely unknown. In recent years, candidates for DM particles have been sought through three main strategies: direct detection, indirect detection, and collider searches (for reviews, see e.g., Refs. \cite{Nature_dd,Nature_id,Nature_collider}). Despite some suggestive signals in these searches (see e.g., Refs. \cite{DAMA01,DAMA02,Cuoco2017,Cui2017,Niu2018b,Niu2019a} and references therein), no confirmed evidence has been obtained yet.

As the ultimate evolutionary stage of most stars in our galaxy \cite{Fontaine2001}, white dwarfs (WDs) possess relatively simple interior structures, consisting of an electron-degenerate core and an atmosphere envelope, and are considered to be the most promising laboratories for measuring DM-electron interactions \cite{Isern2022}.

In our galaxy, DM particles inevitably traverse WDs, losing energy upon scattering with the star's constituents (nuclei and electrons). If these DM particles' velocity, post deceleration, falls below the WD's escape velocity, they become captured and gravitationally bound to the star. These captured DM particles may then accumulate, annihilate within the WD, or evaporate from it, thereby disrupting the star's standard evolutionary trajectory as dictated by stellar evolution theory.

Fortunately, for pulsating WDs, both their interior structures and evolutionary rates can be precisely determined through the pulsation periods and their variation rates \cite{Winget2008,Fontaine2008,Althaus2010,Calcaferro2017}. Consequently, DM-related processes within a WD (capture, evaporation, and annihilation) can be calculated, allowing us to predict its evolutionary rates. By comparing these predictions with observations, we can deduce the properties of DM.

{\bf Period Variation Rates of Pulsating WDs.}
The period variation rate of a pulsating WD's pulsation mode (denoted as $\dP \equiv \diff P/\diff t$), is generally described by the equation \cite{Winget1983}:
\begin{equation}
  \label{eq:WD_rate_R}
  \frac{\dP}{P} \simeq -\frac{1}{2} \frac{\dot{T}_{c}}{T_{c}} + \frac{\dot{R}_{*}}{\Rstar},
\end{equation}
where $P$ signifies the pulsation period, $T_{c}$ is the core temperature of the WD, and $\Rstar$ represents the WD's radius. Utilizing the mass-radius relationship characteristic of low-mass WDs ($\Rstar \propto \Mstar^{-\frac{1}{3}}$), the equation can be reformulated as:
\begin{equation}
  \label{eq:WD_rate_M}
  \frac{\dP}{P} \simeq -\frac{1}{2} \frac{\dot{T}_{c}}{T_{c}} - \frac{1}{3}\frac{\dot{M}_{*}}{\Mstar},
\end{equation}
with $\Mstar$ being the WD's mass.

According to Eq. (\ref{eq:WD_rate_M}), the impact of DM-related processes on the period variation rate $\dP$ in pulsating WDs is evident \cite{Niu2024_DM01}\footnote{In this analysis, we consider $P$, $T_{c}$, and $\Mstar$ as constants.}: (a) Capture and accumulation of DM would lead to a reduction in $\dP$, as DM particles impart kinetic energy to the star's material  ($\dot{T}_{c} > 0$) and enhance the star's mass ($\dot{M}_{*} >0$); (b) Evaporation of DM results in an increase in $\dP$, with the star's material transferring kinetic energy to DM particles ($\dot{T}_{c} < 0$) and the star's mass diminishing ($\dot{M}_{*} <0$); (c) Annihilation of DM decreases $\dP$, as DM particles introduce energy into the star ($\dot{T}_{c} > 0$) (see for e.g., \cite{Niu2018_dav}).

The intriguing aspect arises from contrasting the observed period variation rates with the predictions of stellar evolution for certain pulsating WDs. To date, the period variation rates for several pulsating WDs, namely G117-B15A, R548, L19-2, and PG 1351+489, have been ascertained through extensive time-series photometric observations. Yet, the observed period variation rates for the stable pulsation modes of these WDs consistently exceed the predictions of stellar evolution theory (see Table \ref{tab:WDs}), suggesting a more rapid cooling progression that might be attributed to an additional cooling mechanism, such as the presence of axions \cite{Corsico2016,Romero2012,Sullivan2015,Battich2016}.

\begin{table*}[htp!]
  \centering
  \caption{Information of the Four Pulsating White Dwarfs.}
  \label{tab:WDs}
  \begin{ruledtabular}
  \begin{tabular}{l|cccc}
ID    &{G117-B15A}                   & {R548}                   & {L19-2} & {PG 1351+489} \\
Marks    &{DAV1}                   & {DAV2}                   & {DAV3} & {DBV} \\
    \hline
    $\Pobs\ (\s)$                  & 215.20 & 212.95 & 113.8 & 489.33\\
    $\Pthe\ (\s)$                  & 215.215 & 213.401 & 113.41 & 489.47\\
    ${\dPobs/\Pobs}$ $(\s/\s)$ & $(5.12\pm0.82)\times10^{-15}$ &$(3.3\pm1.1)\times10^{-15}$ & $(3.0\pm0.6)\times10^{-15}$ & $(2.0\pm0.9)\times10^{-13}$ \\
    ${\dPthe/\Pthe}$ $(\s/\s)$ & $1.25\times10^{-15}$ &$1.08\times10^{-15}$ & $1.42\times10^{-15}$ & $0.81\times10^{-13}$\\
    $\Mstar/\Msun$                        & $0.593\pm0.007$ & $0.609\pm0.012$ & $0.705\pm0.023$ & $0.664\pm0.013$ \\
    $\log{(\Lstar/\Lsun)}$         & $-2.497\pm0.030$ & $-2.594\pm0.025$ & $-2.622\pm0.046$ & $-1.244\pm0.030$\\
    $\log{(\Rstar/\Rsun)}$         & $-1.882\pm0.029$ & $-1.904\pm0.015$ & $-1.945\pm0.037$ & $-1.912\pm0.015$\\
    Distance* ($\mathrm{pc}$)   & 57.37& 32.71  & 20.87 & 175.47 \\
\hline
Refs.& \cite{Romero2012,Kepler2021,Bailer2021}& \cite{Romero2012,Mukadam2013,Kepler2021,Bailer2021}& \cite{Sullivan2015,Corsico2016,Pajdosz1995,Bailer2021}& \cite{Corsico2014,Battich2016,Bailer2021} \\
  \end{tabular}
\end{ruledtabular}
\\
\footnotesize{Note: $\Pobs$ is the period of a specific pulsation mode from observation, $\dPobs$ is its variation rate;  $\Pthe$ is the period of a specific pulsation mode from stellar evolution theory, $\dPthe$ is its variation rate; $\Mstar$, $\Msun$ are the mass of the WD and Sun; $\Lstar$, $\Lsun$ are the luminosity of the WD and Sun; $\Rstar$, $\Rsun$ are the radius of the WD and Sun; the distances of the WDs to the Sun are obtained based on Gaia DR3 \cite{Bailer2021}.}
\end{table*}

In this study, we propose that the supplementary cooling mechanism is rooted in effective evaporation, triggered by the elastic scattering between DM particles and the WD's constituents. While we cannot yet provide conclusive evidence for the existence of DM particles based on the observed period variation rates (see, e.g., \cite{Niu2024_DM01}), we can cautiously rule out certain DM parameter spaces and establish constraints on the properties of DM particles.

{\bf WDs' Cooling by DM Evaporation.}
As galactic DM particles traverse a WD, some inevitably lose energy and become captured by the star. Concurrently, these captured DM particles, having gained sufficient energy, are released back into space through evaporation.\footnote{Here, we focus on DM particles that do not annihilate.}
The temporal evolution of the total number of DM particles within the star, denoted as $\Ndm$, is described by:
\begin{equation}
  \frac{\diff \Ndm}{\diff t} = C_{*} - E_{*} \cdot \Ndm,
\label{eq:Ndm}
\end{equation}
where $C_{*}$ represents the star's DM particle capture rate, $E_{*}$ signifies the DM particle evaporation rate. The solution to Eq. (\ref{eq:Ndm}) is given by:
\begin{equation}
  \Ndm(t) = C_{*} t \cdot \left(\frac{1-e^{-E_{*}t}}{E_{*}t}\right).
\label{eq:sol_cap_eva}
\end{equation}

The equilibrium between capture and evaporation is examined on a timescale much shorter than that of stellar evolution. At equilibrium, the star captures and evaporates DM particles at a rate of $C_{*}$ per unit time. These particles facilitate a novel pathway for energy transfer between the star and its surroundings, altering the star's conventional cooling process.

Within a WD, the capture and evaporation rates for nuclei and electrons, represented by $C_{*}$ and $E_{*}$ respectively, differ due to their distinct interaction cross sections and the state of matter, with the majority of electrons existing in a Fermi degenerate state. The calculations for $C_{*}$ and $E_{*}$ for both nuclei and electrons are detailed in the Supplementary Material.

In a WD at equilibrium regarding DM capture and evaporation, the capture process transfers DM energy to the star, denoted as $\Ein$, whereas the evaporation process transfers energy from the star to the DM and subsequently to the external environment, denoted as $\Eout$. Typically, these two processes result in unequal energy exchanges, leading to a net energy change ($\Enet \equiv \Eout - \Ein$) for the WD at equilibrium. Observations indicate that $\Enet > 0$, meaning the WDs experience a net energy loss. The specific expressions for $\Ein$ and $\Eout$ for both nuclei and electrons are provided in the Supplementary Material.

Given that electrons are significantly more efficient in mediating energy exchanges between DM particles and the WD at equilibrium than nuclei (for further insights, refer to \cite{Niu2024_DM01}), we disregard the capture and evaporation of DM by nuclei in our subsequent analysis.

In the equilibrium state, the number of DM particles within a WD remains constant at $C_{*}/E_{*}$, implying $\dot{M}_{*} = 0$. According to Eq. (\ref{eq:WD_rate_M}), if $\Enet > 0$ (which results in $\dot{T}_{c}<0$), an increased $\dot{P}$ is obtained, aligning with the observed outcomes for WDs. In such a scenario, the net energy transferred can be viewed as an alternative form of luminosity, emanating from DM particles ($\Ldm \equiv \Enet$).

Drawing a parallel to the case of axions \cite{Isern1992,Corsico2001}, the relationship is expressed as:
\begin{equation}
  \label{eq:dPdL}
  \frac{\dPobs}{\dPthe} = \frac{\Lstar+\Ldm}{\Lstar},
\end{equation}
where $\dPthe$ is the period variation rate as predicted by stellar evolution theory; $\dPobs$ is the observed period variation rate; and $\Lstar$ is the luminosity derived from asteroseismology models.

{\bf Results and Discussions.}
Utilizing the observed period variation rates from the four WDs, we employ Eq. (\ref{eq:dPdL}) to independently delineate the excluded regions of the DM-electron interaction parameter space. As depicted in Figure \ref{fig:sigma_mass}, we present the $95\%$ excluded lines for two scenarios of the DM form factor, namely $F(q)=1$ and $F(q) = (\alpha m_{e})^{2}/q^{2}$. The bands in the figure signify the typical uncertainties associated with $\dPthe$, stemming from theoretical predictions which range from $\sim 9\% - 15\%$ \cite{Kepler2021}. In this study, we have prudently selected a $15\%$ uncertainty to ensure a conservative approach.

\begin{figure}[htp!]
  \centering
  \includegraphics[width=0.49\textwidth]{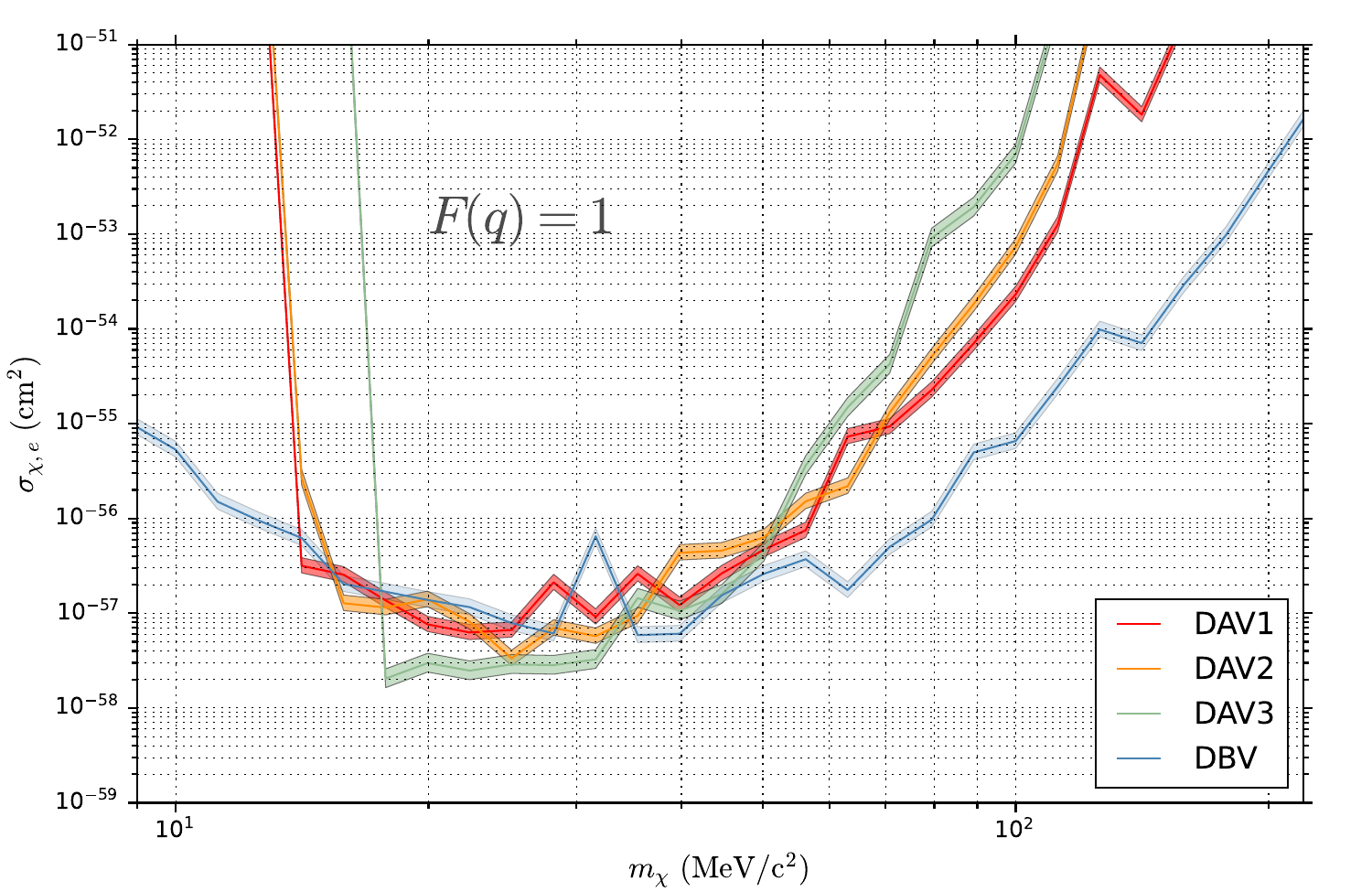}
  \includegraphics[width=0.49\textwidth]{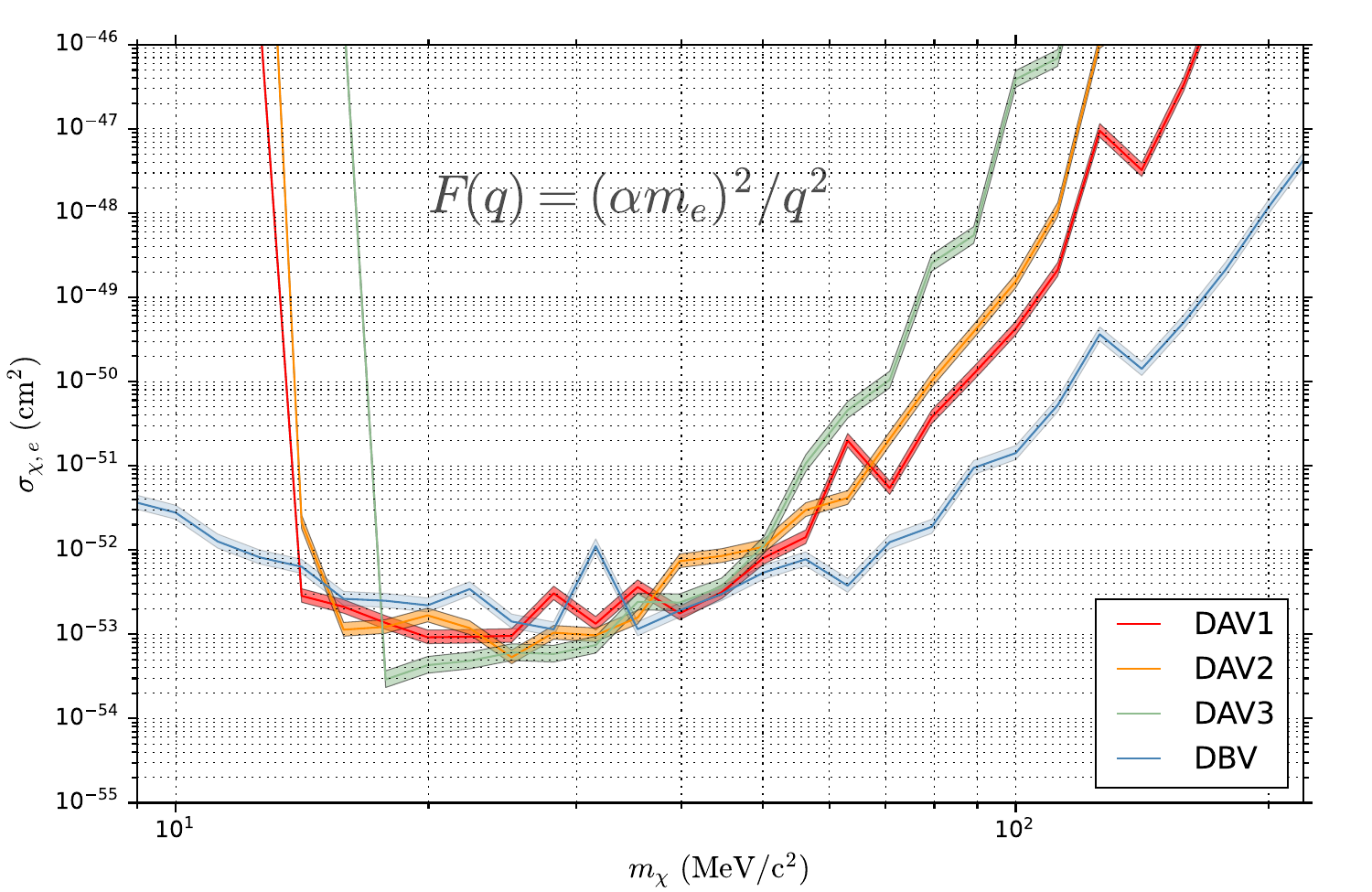}
  \caption{$95\%$ C.L. upper limits on DM-electron interactions as inferred from the four WDs. The scenarios for $F(q) = 1$ (top) and $F(q) = (\alpha m_{e})^{2}/q^{2}$ (bottom) are both considered. The bands denote the uncertainties originating from $\dPthe$.}
  \label{fig:sigma_mass}
\end{figure}

The refined constraints on the DM-electron interactions are articulated as follows:
\begin{itemize}
  \item For $F(q) = 1$, in the DM mass range $20 \MeV/c^{2} \lesim \mdm \lesim 80 \MeV/c^{2}$ with a cross-section limit of $\sige \lesim 10^{-56} \cm^{2}$;
  \item For $F(q) = (\alpha m_{e})^{2}/q^{2}$, in the DM mass range $20 \MeV/c^{2} \lesim \mdm \lesim 70 \MeV/c^{2}$ with a limit of $\sige \lesim 10^{-52} \cm^{2}$.
\end{itemize}

These constraints significantly surpass current direct detection limits, offering over fifteen orders of magnitude improvement in the $10-200 \MeV/c^{2}$ mass range \cite{XENON1T2022,PandaX-II2021,DAMIC-M2023,DarkSide-502023,SENSEI2023,PandaX-4T2023,CDEX2024,DAMIC-M2024,Zema2024}. This advancement is attributed to the unique and advantageous configuration of WDs, which facilitates an efficient capture and evaporation of DM particles within specific mass intervals through electron interactions.

In WDs with core temperatures around $10^{7} \mathrm{K}$ and masses in the $0.5-0.7 \Msun$ range, characteristic of those in this study, DM particles captured by the WD's gravity, within the $10-200 \MeV/c^{2}$ mass range, possess kinetic energies of $\mdm \vesc^{2}/2$. This energy is exactly between the thermal energy ($\sim kT_{c}$) and the Fermi degenerate energy ($\sim \mu_{F}$) of the electrons, rendering such WDs as 'resonant cavities' for DM capture and evaporation via electrons.

Furthermore, the neutrino fog, primarily originating from solar neutrinos and typically considered a theoretical lower limit in direct DM detection experiments \cite{Neutrino2024}, is naturally circumvented in this scenario.

The investigative approach presented in this work may offer valuable insights for DM particle searches. It hinges on the necessity to accurately determine the period variation rates ($\dPobs$) of certain stable pulsation modes in pulsating WDs, thereby ascertaining their evolutionary rates. This, however, is not a straightforward task, necessitating prolonged time-series photometric observations that span decades.

Fortunately, with the growing repository of time-series photometric data from pulsating WDs, especially from space-based telescopes such as TESS, the period variation rates will be determined with increasing precision. Each dataset has the potential to provide valuable clues regarding DM particles. At the very least, it will establish an upper limit in the parameter space, contingent upon the specific properties of the WD.

We advocate for the proposal and implementation of additional innovative scenarios to perform cross-checks on the findings of this research.

\begin{acknowledgments}
  J.S.N. acknowledges support from the National Natural Science Foundation of China (NSFC) (No. 12005124) and the Applied Basic Research Programs of Natural Science Foundation of Shanxi Province (No. 202103021223320).
\end{acknowledgments}

%

\clearpage
\newpage
\maketitle
\onecolumngrid

\begin{center}
\textbf{\Large Supplementary Material}\\
\vspace{0.1in}
\textbf{\large Exploring the Dark Frontier: White Dwarf-Based Constraints on Light Dark Matter} \\ 
\vspace{0.05in}
{}
{Jia-Shu Niu}

\end{center}

In this supplementary section, we provide an exhaustive account of the capture and evaporation rates for DM particles, denoted as $C_{*}$ and $E_{*}$, respectively, as well as the energy transferred during these processes, $\Ein$ and $\Eout$.

\section{Capture and Evaporation rates of DM Particles}
\label{app:01}

Generally, there exists an upper bound on the capture rate of a WD for DM particles, defined under the scenario where every DM particle passing through the star is captured. This condition of saturated capture is referred to as the geometric limit, expressed by the equation \cite{Bell2021}:
\begin{equation}
\label{eq:cap_geom}
C_\mathrm{geo}= \frac{\pi \Rstar^{2}}{3\vstar} \frac{\rhodm}{\mdm} \left[ (3 \vesc^{2}(\Rstar)+3\vstar^{2}+v_{d}^{2}) \cdot \mathrm{Erf} \left(\sqrt{\frac{3}{2}} \frac{\vstar}{v_{d}} \right) + \sqrt{\frac{6}{\pi}} \vstar v_{d} \cdot \exp \left( - \frac{3\vstar^{2}}{2v_{d}^{2}} \right) \right].
\end{equation}
Here, the local DM density around the Sun is denoted by $\rhodm = 0.3\ \GeV/\cm^{3}$; $\mdm$ represents the mass of the DM particle; $\Rstar$ is the star's radius; $\vesc (r)$ is the escape velocity of a DM particle at a distance $r$ from the star's center; $\vstar = 220\ \km/\s$ is the Sun's relative velocity with respect to the DM frame where the average DM speed is null; and $v_d = 270 \ \km/\s$ is the dispersion velocity of DM particles in the vicinity of the Sun. Given the relatively small distances from the Sun to the four WDs compared to the galactic center ($\sim 8.5 \kpc$), the values of $\rhodm$, $\vstar$, and $v_d$ are considered constant throughout this study.

In the non-saturation capture condition, the capture rate of DM particles is formulated as follows \cite{Garani2019,Bell2021}:
\begin{equation}
  \label{eq:cap_non}
  C_\mathrm{ngeo}= \int_{0}^{\Rstar} \frac{\rhodm}{\mdm} 4\pi r^{2} \diff r \int_{0}^{\infty} \frac{f_{\vstar}(\udm)}{\udm} w(r) \diff \udm \int_{0}^{\vesc} R^{-}(w \rightarrow v) \diff v,
\end{equation}
where $\udm$ represents the velocity of DM particles at a significant distance from the star; $w(r) = \sqrt{\udm^{2} + \vesc^{2}(r)}$ signifies the velocity of a DM particle at an arbitrary distance from the star's center; $R^{-}(w \rightarrow v)$ denotes the differential scattering rate for capture, considering a DM particle with initial velocity $w$ scattering to a lower velocity $v$ ($w > v$), a process that is contingent upon the star's material equation of state and varies between nuclei and electrons; $f_{\vstar}(\udm)$ is the DM velocity distribution, typically characterized by a Maxwell-Boltzmann distribution \cite{Busoni2017}:
\begin{equation}
\label{eq:fDM_out}
f_{\vstar}(\udm) = \frac{\udm}{\vstar} \sqrt{\frac{3}{2\pi (v_{d}^{2} + 3kT_{*}/m_{T})}} \left( \exp \left[- \frac{3(\udm - \vstar)^{2}}{2(v_{d}^{2} + 3kT_{*}/m_{T})} \right] -  \exp \left[- \frac{3(\udm + \vstar)^{2}}{2(v_{d}^{2} + 3kT_{*}/m_{T})} \right] \right),
\end{equation}
In this context, $k$ symbolizes the Boltzmann constant; $T_{*}$ refers to the temperature of the star, which is generally radius-dependent; $m_{T}$ is the mass of the target particles, which could be either nuclei or electrons.

The capture rate is generally defined as the minimum of the geometric and non-geometric capture rates:
\begin{equation}
  \label{eq:cap}
  C_\mathrm{*}= \min \{C_\mathrm{geo}, C_\mathrm{ngeo}\}.
\end{equation}

The evaporation rate is articulated with precision in the following equation \cite{Gould1987b,Garani2017,Busoni2017}:
\begin{equation}
  \label{eq:eva}
  E_{*}= \int_{0}^{\Rstar} \ndm (r) 4\pi r^{2} \diff r \int_{0}^{\vesc} f_{\chi}(w,r) 4\pi w^{2} \diff w \int_{\vesc}^{\infty} R^{+}(w \rightarrow v) \diff v,
\end{equation}
where $R^{+}(w \rightarrow v)$ represents the differential scattering rate for evaporation, contingent upon the target particles within the star; $\ndm (r)$ denotes the normalized radial distribution of DM \cite{Gould1987b,Garani2017,Garani2019}:
\begin{equation}
  \label{eq:DM_dis}
  \ndm (r) = \frac{4}{\rdm^{3} \sqrt{\pi}} \exp \left(- \frac{r^2}{\rdm^{2}} \right),\text{with}\  \rdm = \sqrt{\frac{3kT_{*}}{2 \pi G \rhostar \mdm}},
\end{equation}
$G$ being the gravitational constant, and $\rhostar$ symbolizing the star's density; $f_{\chi}(w,r)$ is the velocity distribution of the thermalized DM particles, conforming to a Maxwell-Boltzmann distribution truncated at the escape velocity $\vesc(r)$ as delineated by \cite{Gould1987b,Garani2017,Garani2019}:
\begin{equation}
\label{eq:fDM_out}
f_{\chi}(w,r) = \frac{1}{\pi^{3/2}} \left(\frac{\mdm}{2kT_{*}}\right)^{3/2} \frac{\exp \left(-\frac{\mdm w^{2}}{2kT_{*}} \right) \Theta(\vesc(r) - w)}{\mathrm{Erf} \left(\sqrt{\frac{\mdm \vesc^{2}(r)}{2kT_{*}}} \right) - \frac{2}{\sqrt{\pi}} \sqrt{\frac{\mdm \vesc^{2}(r)}{2kT_{*}}} \exp \left(-\frac{\mdm \vesc^{2}(r)}{2kT_{*}} \right)}.
\end{equation}

In the domain of nuclear interactions, the differential scattering rates for capture and evaporation are defined as follows \cite{Busoni2017}:
\begin{equation}
  \label{eq:R_cap_n}
  R^{-}(w \rightarrow v)= \frac{32\mu_{+}^{4}}{\sqrt{\pi}} \kappa^{3} n_{n}(r) \frac{\diff \sigma_{N}}{\diff \cos \theta} \frac{v}{w} \int_{0}^{\infty} \diff \vs \int_{0}^{\infty}  \vt e^{-\kappa^{2} v_{T}^{2}} H^{-}(\vs,\vt,w,v) \diff \vt,
\end{equation}
and
\begin{equation}
  \label{eq:R_eva_n}
  R^{+}(w \rightarrow v)= \frac{32\mu_{+}^{4}}{\sqrt{\pi}} \kappa^{3} n_{n}(r) \frac{\diff \sigma_{N}}{\diff \cos \theta} \frac{v}{w} \int_{0}^{\infty} \diff \vs \int_{0}^{\infty}  \vt e^{-\kappa^{2} v_{T}^{2}} H^{+}(\vs,\vt,w,v) \diff \vt.
\end{equation}
The associated parameters are detailed as:
\begin{align}
  \label{eq:notion_n}
  \begin{aligned}
  \mu_{\pm} &= \frac{\mu \pm 1}{2}, \mu = \frac{\mdm}{m_{n}}, \kappa^{2} = \frac{m_{n}}{2kT_{*}},\\
  v_{T}^{2} &= 2\mu \mu_{+} \vt^{2} + 2\mu \vs^{2} - \mu w^{2}, \\
  H^{-}(\vs,\vt,w,v) &= \Theta(\vt+\vs-w) \Theta(v-|\vt-\vs|), \\
  H^{+}(\vs,\vt,w,v) &= \Theta(\vt+\vs-v) \Theta(w-|\vt-\vs|),
  \end{aligned}
\end{align}
In these expressions, $n_{n}(r)$ signifies the number density of the nucleus; $\vs$ denotes the velocity in the star's frame of the center of mass (CM) of the scattering event, while $\vt$ is the velocity of the DM in the CM frame; $m_{n}$ is the mass of the nucleus.

For electrons, the differential scattering rates for capture and evaporation can be expressed as \cite{Garani2019}
\begin{equation}
  \label{eq:R_cap_e}
  R^{-}(w \rightarrow v)= 8 \mu_{+}^{4} \sige F(q) n_{e}(r) \frac{v}{w} \int_{0}^{\infty} \diff \vs \int_{0}^{\infty}  \vt f_{p}(E_{p},r) (1-f_{p'}(E_{p'},r)) H^{-}(\vs,\vt,w,v) \diff \vt,
\end{equation}
and 
\begin{equation}
  \label{eq:R_eva_e}
  R^{+}(w \rightarrow v)= 8 \mu_{+}^{4} \sige F(q) n_{e}(r) \frac{v}{w} \int_{0}^{\infty} \diff \vs \int_{0}^{\infty}  \vt f_{p}(E_{p},r) (1-f_{p'}(E_{p'},r)) H^{+}(\vs,\vt,w,v) \diff \vt,
\end{equation}
where the parameters are defined as:
\begin{align}
  \label{eq:notion_e}
  \begin{aligned}
    \mu_{\pm} &= \frac{\mu \pm 1}{2}, \mu = \frac{\mdm}{m_{e}}\\
    f_{p}(E_{p},r) &= \left( \exp \left(\frac{E_{p}-\mu_{F}(r)}{T_{*}} \right) +1 \right)^{-1},\\
    1-f_{p'}(E_{p'},r) &= 1- \left( \exp \left(\frac{E_{p'}-\mu_{F}(r)}{T_{*}} \right) +1 \right)^{-1},\\
    E_{p} &= \frac{1}{2} m_{e} (2\mu \mu_{+} \vt^{2} + 2\mu \vs^{2} - \mu w^{2}),\\
    E_{p'} &= \frac{1}{2} m_{e} (2\mu \mu_{+} \vt^{2} + 2\mu \vs^{2} - \mu v^{2}),
  \end{aligned}
\end{align}
In these expressions, $n_{e}(r)$ signifies the number density of electrons; $m_{e}$ is the electron's mass; $f_{p}(E_{p},r)$ and $1-f_{p'}(E_{p'},r)$ represent the Fermi-Dirac distribution for the electron in its initial and final states, respectively; $\mu_{F}(r)$ is the chemical potential of the electron at position $r$; $H^{\pm}(\vs,\vt,w,v)$ is consistent with the expression given in Eq. (\ref{eq:notion_n}); $F(q)$ is the DM form factor dependent on momentum.

For the scope of this research, two scenarios are considered for the DM form factor: $F(q) = 1$ and $F(q) = (\alpha m_{e})^{2}/q^{2}$, where $\alpha$ is the fine-structure constant and $q$ denotes the transferred momentum.

Additionally, several simplifying assumptions have been applied in our analysis:
(i) We assume a uniform matter distribution within a WD, expressed as $\rhostar (r) = \rhostar = \Mstar/\Vstar$, where $\rhostar$, $\Mstar$ and $\Vstar$ represent the density, mass, and volume of a WD, respectively.
(ii) Given that all four WDs considered in this study are carbon-oxygen core WDs, we adopt a uniform chemical composition across the scattering volume $\Vstar$, utilizing an average atomic weight of 14 for the nucleus \cite{Mukadam2013}.
(iii) We also assume a uniform temperature profile $T_{*}$ within a WD, which is independent of the radial distance $r$, due to the exceptionally high thermal conductivity of the electron-degenerate core.
(iv) Considering that WDs are electrically neutral, we calculate the total number of electrons $N_{e}$ and the number density of electrons $n_{e}$ in a WD using the expressions ${\Mstar}/{2m_{p}}$ and ${\rhostar}/2{m_{p}}$, respectively, where $m_{p}$ is the mass of a proton.

\section{Energy Transferred by DM Capture and Evaporation}
\label{app:02}
The kinetic energy transferred to the star's nuclei or electrons during the capture process is articulated by the following expression, applicable when $C_\mathrm{geo} < C_\mathrm{ngeo}$:
\begin{align}
  \label{eq:Ein}
  \Ein &= \frac{C_\mathrm{geo}}{C_\mathrm{ngeo}} \int_{0}^{\Rstar} \frac{\rhodm}{\mdm} 4\pi r^{2} \diff r \int_{0}^{\infty} \frac{f_{\vstar}(\udm)}{\udm} w(r) \diff \udm \int_{0}^{\vesc} \frac{1}{2} \mdm (w^2 - v^2) \cdot R^{-}(w \rightarrow v) \diff v.
\end{align}

For scenarios when $C_\mathrm{geo} \ge C_\mathrm{ngeo}$, the energy transferred during capture is given by:
\begin{align}
  \Ein &= \int_{0}^{\Rstar} \frac{\rhodm}{\mdm} 4\pi r^{2} \diff r \int_{0}^{\infty} \frac{f_{\vstar}(\udm)}{\udm} w(r) \diff \udm \int_{0}^{\vesc} \frac{1}{2} \mdm (w^2 - v^2) \cdot R^{-}(w \rightarrow v) \diff v.
\end{align}

The kinetic energy transferred from the WD's constituents to the DM particles during the evaporation process is described by:
\begin{equation}
  \label{eq:Eout}
  \Eout = C_{*} \int_{0}^{\Rstar} \ndm (r) 4\pi r^{2} \diff r \int_{0}^{\vesc} f_{\chi}(w,r) 4\pi w^{2} \diff w \int_{\vesc}^{\infty} \frac{1}{2} \mdm (v^2 - w^2) \cdot R^{+}(w \rightarrow v) \diff v.
\end{equation}
These expressions are derived from the foundational equations (\ref{eq:R_cap_n}), (\ref{eq:R_eva_n}), (\ref{eq:R_cap_e}), and (\ref{eq:R_eva_e}).

\end{document}